\documentclass[preprint,NumberedRefs]{IEEEtran}

\pdfoutput=1

\usepackage[T1]{fontenc}
\usepackage[utf8]{inputenc}
\usepackage{textcomp} 

\usepackage{amsmath,amsfonts,amssymb}
\usepackage{algorithmic}
\usepackage{array}
\usepackage[caption=false,font=normalsize,labelfont=sf,textfont=sf]{subfig}
\usepackage{textcomp}
\usepackage{stfloats}
\usepackage{url}
\usepackage{verbatim}
\usepackage{graphicx}

\usepackage[hidelinks]{hyperref}
\usepackage[numbers]{natbib}
\def\BibTeX{{\rm B\kern-.05em{\sc i\kern-.025em b}\kern-.08em
    T\kern-.1667em\lower.7ex\hbox{E}\kern-.125emX}}
\usepackage{balance}

\usepackage{booktabs}
\usepackage{longtable}
\usepackage{calc} 

\usepackage{url}
\graphicspath{{imgs/}}
\makeatletter
\@ifpackageloaded{subfig}{}{\usepackage{subfig}}
\@ifpackageloaded{caption}{}{\usepackage{caption}}
\AtBeginDocument{%
\renewcommand*\figurename{Fig.}
\renewcommand*\tablename{Table}
}
\AtBeginDocument{%
\renewcommand*\listfigurename{List of Figures}
\renewcommand*\listtablename{List of Tables}
}
\newcounter{pandoccrossref@subfigures@footnote@counter}
{\end{figure}%
\addtocounter{footnote}{-\value{pandoccrossref@subfigures@footnote@counter}}
\@for\f:=\global@pandoccrossref@subfigures@footnotes\do{\stepcounter{footnote}\footnotetext{\f}}%
\gdef\global@pandoccrossref@subfigures@footnotes{}}
\@ifpackageloaded{float}{}{\usepackage{float}}
\floatstyle{ruled}
\@ifundefined{c@chapter}{\newfloat{codelisting}{h}{lop}}{\newfloat{codelisting}{h}{lop}[chapter]}
\floatname{codelisting}{Listing}

\makeatother

\def\[#1\]{%
  \begin{equation}#1\end{equation}%
}

\date{\today}

\begin{document}

\title{Comparison of fundamental frequency estimators with subharmonic
voice signals}

\author{
      Takeshi Ikuma, Melda Kunduk, and     Andrew J. McWhorter
\thanks{T. Ikuma (tikuma@ieee.org) is with Department of
Otolaryngology--Head and Neck Surgery, Louisiana State University Health
Sciences Center, New Orleans, LA.}
\thanks{M. Kunduk is with Department of Communication Disorders,
Louisiana State University, Baton Rouge, LA.}
\thanks{A. J. McWhorter is with Department of Otolaryngology--Head and
Neck Surgery, Louisiana State University Health Sciences Center, New
Orleans, LA.}
}

\markboth{Ikuma et al.~- Jan.~2025}%
{Fo estimator comparison with subharmonic voices}

\maketitle

\begin{abstract}
In clinical voice signal analysis, mishandling of subharmonic voicing
may cause an acoustic parameter to signal false negatives. As such, the
ability of a fundamental frequency estimator to identify speaking
fundamental frequency is critical. This paper presents a sustained-vowel
study, which used a quality-of-estimate classification to identify
subharmonic errors and subharmonics-to-harmonics ratio (SHR) to measure
the strength of subharmonic voicing. Five estimators were studied with a
sustained vowel dataset: Praat, YAAPT, Harvest, CREPE, and FCN-F0.
FCN-F0, a deep-learning model, performed the best both in overall
accuracy and in correctly resolving subharmonic signals. CREPE and
Harvest are also highly capable estimators for sustained vowel analysis.
\end{abstract}

\begin{IEEEkeywords}
Disordered voice, Acoustic analysis, Pitch, Subharmonics
\end{IEEEkeywords}

\section{Introduction}\label{introduction}

One of the hallmarks in the pathological voice is the frequent
occurrences of subharmonic phonation
\citep{behrman1998, cavalli1999, kramer2013, ikuma2023a}, causing the
voice to have a rough perceptual quality \citep{omori1997}. Normally,
vocal folds oscillate in unison, locked to the same frequency.
Subharmonic oscillation causes the glottal cycles to have cyclically
varying magnitude, duration, or shape, and voice signals capturing this
behavior are categorized as the type 2 voice signals \citep{titze1994}.

Subharmonic vocal fold vibration may occur in three ways or a
combination thereof. The entire glottal structure may cyclically
modulate with the same cycle
period\citep{kiritani1993, kniesburges2016, ikuma2016a}. Alternately,
two parts of glottis---e.g., left vs.~right vocal folds or anterior
vs.~posterior---may vibrate at different cycle periods (biphonation) but
in a synchronized manner
\citep{kiritani1993, mergell2000, neubauer2001, ikuma2016a}.
Synchronization imposes these cycles to align periodically. Finally,
subharmonic voicing may also occur with additional vibration of the
ventricular or aryepiglottic folds \citep{titze2024}. All these forms of
irregular vocal fold vibration produce acoustic signals with subharmonic
components or additional tones that relate to the harmonic tones by
rational frequency ratios.

An important trait of subharmonic voice signals is that they are nearly
periodic, i.e., they share the same basic quality with normal voice
signals. The fundamental period of a subharmonic signal, however, is
longer than the normal as its period spans over multiple glottal cycles.
When normal harmonic oscillation with \(T\)-second period bifurcates to
period-\(M\) subharmonic oscillation, the true period of the signal
elongates to \(M \times T\) seconds. Here, a positive integer \(M\) is
the subharmonic period in glottal cycles. In the case of the subharmonic
biphonation with two glottal fundamental periods \(T_1\) and \(T_2\),
there exist two subharmonic periods \(M_1\) and \(M_2\) such that
\(M_1T_1 = M_2T_2\), which constitutes the common true period.

Importantly, the vast majority of clinical acoustic parameters---such as
jitter, shimmer, and harmonics-to-noise ratio---rely on the knowledge of
the speaking fundamental frequency \(f_o = 1/T\) \citep{kaypentax2008}.
Thus, the effectiveness of these parameters hinges on the accuracy of
the estimated \(f_o\), but the near-periodicity of subharmonic signals
often steers a fundamental frequency estimator away from \(f_o\), either
detecting \(f_o/M\) or reporting unvoiced. While reporting unvoiced
merely declares that most of the acoustic parameters cannot be computed
for the signal segment, misdetecting \(f_o/M\) as \(f_o\) has
potentially negative consequences. The resulting measurements may fall
into the normal parameter ranges by treating the subharmonic tones as
the harmonic tones, thereby underreporting the severity.

Many of the \(f_o\) estimation algorithms (also known as pitch detection
algorithms), however, are designed to detect the true fundamental
frequency of their input signals rather than the speaking \(f_o\) based
on the assertion that the voice is a nearly periodic phenomenon. This
assertion is appropriate for normal speech processing tasks because
subharmonic phonation is rare in the vocally healthy population although
it can be voluntarily produced as vocal fry \citep{hollien1966} or as a
singing technique \citep{herbst2017}.

The knowledge of subharmonics in voice and its implication for
pathological voice analysis have led to several \(f_o\) detection
algorithms to account for this possibility. In 1990, Hedelin and
Huber\citep{hedelin1990} proposed to use the amplitude compression
technique among others to reduce the abrupt amplitude change associated
with irregular voicing. Sun \citep{sun2000} proposed an algorithm to
estimate \(f_o\) and detect the period-doubling subharmonics
simultaneously based on the subharmonics-to-harmonics ratios. However,
extending this method to support higher subharmonic period is not
trivial despite that higher-period subharmonics are known to be present
in pathological voice \citep{ikuma2023a}. \citet{hlavnicka2019} improved
Sun's method with more elaborate \(f_o\) estimation using a Kalman
filter although still limited to the period-doubling subharmonics.
Hagmüller and Kubin\citep{hagmuller2006} proposed the use of the phase
space from the dynamical systems analysis with a mixed result, targeting
the subharmonic voice. Aichinger et al.
\citep{aichinger2015, aichinger2017b, aichinger2018a} proposed to use
parametric models with multiple harmonic sources to resolve two
oscillators of biphonation cases. None of these algorithms, however,
have yet been adopted by voice and speech analysis tools.

Existing \(f_o\) estimators have been evaluated for their handling of
pathological voices
\citep{laver1982, parsa1999, jang2007, tsanas2014, vaysse2022}. To
assess the effect of the subharmonics, Jang et al.\citep{jang2007} and
Vaysse et al.\citep{vaysse2022} quantified the \(f_o\)-halving error
rate. It is a generic coarse metric to identify excessive
under-estimation errors rather than classifying whether an estimator
selected \(f_o/2\) or \(f_o\). The prior studies also only reported
per-recording averaged results rather than individual estimates.
Averaging obfuscates the impact of subharmonics because pathological
voice can be highly volatile; its mode of operation can change
intermittently, and even rapidly
\citep{dejonckere1983, titze1994, behrman1998, ikuma2023a}.

The current study aimed to assess the performance of five selected
modern \(f_o\) estimators with a reference autocorrelation-based
estimator. The study focused on their per-frame handling of subharmonic
sustained vowel samples. A quality-of-estimate classification method was
devised to identify possible undesirable detections of subharmonic
fundamental frequency over other types of estimation errors. Moreover,
the subharmonics-to-harmonics ratio (SHR) measurements were used to gain
further insights.

\section{Methods}\label{sec:methods}

\textbf{Acoustic Data.} Sustained /a/ audio data of KayPENTAX Disordered
Voice Database \citep{kaypentax2006} was used in the study. All the
available recordings (710) were considered, including those without
demographics and diagnostic information. This database supplies its data
at two different sampling rates: 53 normal voice recordings at 50000
samples/second (S/s) and 657 pathological voice recordings at 25000 S/s.
The recordings come pre-trimmed without voice onsets and offsets. The
normal recordings are three seconds long while the pathological
recordings are variable lengths: most are one-second long with the
shortest of 450 milliseconds (ms). All recordings were resampled to 8000
S/s for the study. Each recording was assessed in a 50-ms interval,
yielding 16174 total intervals.

\textbf{\(f_o\) Annotation.} The database does not provide information
pertaining to the fundamental frequency \(f_o\) or the presence of
subharmonics. As such, the truth \(f_o\) value (denoted by \(f_o^*\)) of
each signal interval was manually annotated in three steps. First, the
initial estimates were gathered from the Praat \footnote{https://github.com/praat/praat/tree/v6.1.38}.
Then, these estimated \(f_o\)'s were reviewed and adjusted manually with
a custom computer program, which superimposes a manually adjustable
\(f_o\) track on a narrowband spectrogram with audio playback
capability. All instances of voice breaks and loss of harmonic structure
were also confirmed to be excluded from the study. Finally, the
estimates were refined using the time-varying harmonic model with a
gradient-based optimization \citep{ikuma2022a}. After the elimination of
unvoiced segments, the analysis sample size reduced to 15941 intervals
out of 703 recordings.

Spectrogram with audio playback was found sufficient to identify the
\(f_o\)'s of normal voicing and those with weak or intermittent
subharmonics. There were a few cases for which decisive judgments could
not be reached due to strong sustained subharmonics. For those cases,
the speaker's information was first checked for sex appropriate \(f_o\)
(e.g., the lowest tone under 100 Hz is likely a subharmonic tone for a
female voice), then the signal waveform was inspected in Praat to
identify the glottal cycles.

\begin{figure}
\centering
\includegraphics[keepaspectratio]{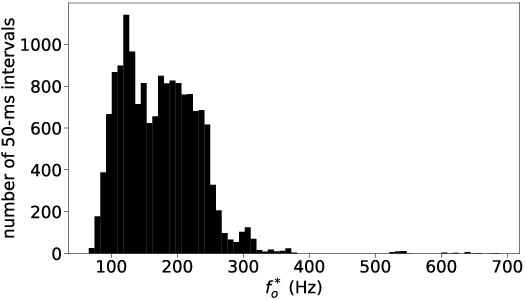}
\caption{Histogram of the annotated fundamental frequencies \(f_o^*\)
(15941 samples).}\label{fig:fo_dist}
\end{figure}

Fig. \ref{fig:fo_dist} illustrates the \(f_o^*\) distribution of the
study dataset. It follows the expected bimodal nature due to the sexual
difference and fully covers the expected frequency range of modal voice
\citep{baken2000}. The observable outliers with abnormally high \(f_o\)
were all included in the study.

\textbf{Detection Algorithms.} Five \(f_o\) detection algorithms were
assessed: Praat \(f_o\) detector\citep{boersma1993}, Harvest
estimator\citep{morise2017}, ``yet another algorithm for pitch
tracking'' (YAAPT)\citep{zahorian2008}, convolutional representation for
pitch estimation (CREPE) \citep{kim2018}, and fully convolutional
network for \(F_0\) estimation (FCN-F0) \citep{ardaillon2019}. These
were chosen by the general popularity (Praat, CREPE) and based on their
reported ability to handle subharmonics\citep{vaysse2022} (YAAPT,
FCN-F0), and the state-of-art successor (Harvest) of another popular
estimator, nearly-defect free (NDF) estimator\citep{kawahara2005}.
Finally, an alternate configuration of the Praat \(f_o\) detector was
used to obtain another set of estimates based only on the
autocorrelation function (ACF) as the baseline.

Publicly available implementations of these algorithms were used in the
assessment with their default parameter settings unless noted otherwise
below. All the estimators, except for CREPE and FCN-F0, were configured
to limit their estimates to be between 60 and 700 Hz based on the
observed \(f_o\) values during the annotation step. The \(f_o\) ranges
of CREPE and FCN-F0 models are fixed by the structures of the models.
The evaluation interval was either changed to 50 ms to match the
annotation interval (Praat, ACF, CREPE, and FCN-F0) or the results from
the middle of the intervals were picked (Harvest and YAAPT).

\emph{Praat} \(f_o\) estimator \citep{boersma1993} is based on the ACF
and uses the Viterbi algorithm\citep{viterbi1967} for postprocessing.
Strong peaks of the ACF are selected as the candidates, each with
computed \(f_o\) and pitch strength estimates at each time-step. The
Viterbi algorithm is a dynamic programming algorithm to obtain the most
likely sequence of \(f_o\) over the time steps by selecting the best
path among the candidates, including the possibility of an unvoiced
time-step. The path is determined by the candidates' pitch strengths and
the cost associated with switching the frequency. The \emph{ACF}
estimator is a reconfigured Praat estimator with the Viterbi
postprocessing stage disabled by setting the \texttt{Octave-jump\ cost}
and \texttt{Voiced\ /\ unvoiced\ cost} parameters to zero
\citep{boersma1993}. This forces the estimator to select the \(f_o\)
candidate with the strongest autocorrelation in each interval. The
Parselmouth Python package \citep{jadoul2018} was used to run Praat for
this study.

The \emph{Harvest} estimator \citep{morise2017} is the latest \(f_o\)
estimator for the WORLD vocoder\citep{morise2016}. It uses a filterbank
to find an \(f_o\) candidate in each filter with logarithmically spaced
center frequency and derives the final \(f_o\) estimates with using a
complex set of conditions such as harmonic reliability, frequency jump,
and voicing duration. The Python implementation\footnote{https://github.com/JeremyCCHsu/Python-Wrapper-for-World-Vocoder/tree/v0.3.4}
was used in this study.

The \emph{YAAPT} estimator\citep{zahorian2008} combines multiple
techniques including the normalized cross-correlation, spectral harmonic
correlation, and dynamic programming. A Python version of this
estimator\footnote{https://github.com/bjbschmitt/AMFM\_decompy/tree/01fe42}
was used for the assessment.

The \emph{CREPE} deep-learning model \citep{kim2018} is a six-layer
convolutional neural network (CNN) model. The dense output layer
produces 360 nodes, representing \(f_o\) between 31.7 and 2005.5 Hz,
spaced logarithmically. Each output node represents the likelihood of
the fundamental frequency in the proximity of its assigned \(f_o\)
value. The final estimate is obtained via weighted averaging. The model
takes 1024-sample input at 16000 S/s; hence the input signal (which were
initially resampled to 8000 S/s) was again resampled up to 16000 S/s.
The study used the supplied pretrained model coefficients, which were
trained with synthesized music data\citep{kim2018}.

The \emph{FCN-F0} deep-learning model \citep{ardaillon2019} is an
extension of the CREPE model by replacing the final dense layer with
another convolution layer, thereby turning the model into a fully
convolutional network, and adjusting its hyperparameters for improved
performance. It uses the input sampling rate of 8000 S/s and produces
the likelihood outputs over a modified frequency range between 30 and
1000 Hz with 486 bins. This model was published with three pretrainined
models, and this study used the FCN-993 model with 993 input frame size.
The model coefficients were trained with English and French
non-pathological speech corpus\citep{ardaillon2019}. Author (TI)'s
repackaged version\footnote{https://github.com/tikuma-lsuhsc/python-ml-pitch-models/releases/tag/v0.0.1}
was used for both FCN-F0 and CREPE.

\textbf{Quality-of-Estimate Classification.} Each estimator's output
(denoted by \(\hat{f}_o\)) for every 50-ms time interval was checked for
its correctness against the annotated truth, \(f_o^*\). An estimate
\(\hat{f}_o\) was labeled either correct, erroneous due to subharmonic
error, or erroneous due to other types of estimation error. The
subharmonic error refers to an event when an estimator output near
\(f_o^*/M\) with some positive integer \(M>1\). This could be caused
either by the true presence of the subharmonics or by an estimator
incidentally picking less than the largest frequency to explain the
periodicity. The former is not technically an error for the sake of
mathematical correctness but is an undesirable outcome for the purpose
of clinical voice analysis.

\begin{figure}
\centering
\includegraphics[keepaspectratio]{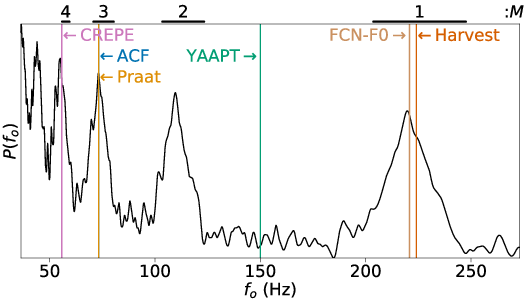}
\caption{(color online) Illustration of harmonic power profile
\(P(f_o)\) and the accuracy classification of \(\hat{f_o}\). The
vertical lines indicate the \(\hat{f_o}\) of the labeled estimator, and
the horizontal bars along the top edge indicate the intervals associated
with the truth (\(M=1\)) and subharmonic errors (\(M>1\))
(\(SHR = -6.8\) dB).}\label{fig:msub_demo}
\end{figure}

Both correct and subharmonic error labels were casted based on the
closeness of \(\hat{f}_o\) to \(f_o^*/M\), \(M>0\). The closeness was
evaluated based on where they are placed on the harmonic power profile,
which is defined by
\begin{equation}{P(f_o) = \sum_{k=1}^{K(f_o)} |S_{xx}(kf_o)|^2,
}\end{equation} where \(S_{xx}(f)\) is the periodogram of the input
signal \(x_n\) with 50-ms Hamming window and \(K(f_o)\) is the maximum
observable number of the harmonics of a signal with the fundamental
frequency \(f_o\), i.e.,
\begin{equation}{K(f_o) = \left\lfloor \frac{f_s}{2 f_o} \right\rfloor.
}\end{equation} Here, \(f_s\) is the sampling rate and
\(\lfloor \cdot \rfloor\) is the floor function. An example of
\(P(f_o)\) is shown in Fig. \ref{fig:msub_demo}.

Given the annotated truth \(f_o^*\), \(f_o\)-intervals were defined for
the correct estimate (\(M=1\)) and subharmonic error estimate (\(M>1\)).
The \(M\)th interval contains \(f_o^*/M\) and is bounded by a pair of
the neighboring local minima of \(P(f_o)\). If \(\hat{f}_o\) fell in one
of the intervals, it was said to be close enough to be correct (\(M=1\))
or was erroneous with subharmonic error (\(M>1\)). If no such \(M\) was
found, \(\hat{f}_o\) was labeled erroneous with some other error. On
Fig. \ref{fig:msub_demo}, these intervals are indicated by the
horizontal black bars at the top of the axes, and two estimators
(Harvest and FCN-F0 under \(M=1\)) yielded correct estimates while CREPE
(\(M=4\)), Praat and ACF (\(M=3\)) were subject to subharmonic errors,
and YAAPT failed with other error,

The \(f_o\)-intervals were numerically evaluated with a peak-picking
algorithm. The periodogram was densely sampled with the discrete Fourier
transform with 0.5-Hz resolution.

\textbf{Analysis Objectives.} The per-frame analyses were conducted to
observe three performance aspects: (1) the mapping between the estimates
and the truths, (2) the detection error rates of the estimators, and (3)
how the input intervals which caused subharmonic errors on the ACF were
handled by the other estimators. Since the ACF estimator is the most
primitive among those tested, it was thought to be most sensitive to the
subharmonics.

In addition to the error rates, the subharmonics-to-harmonics ratios
(SHRs) were estimated with the ACF outcomes with subharmonic errors. The
SHR is a ratio of the power of subharmonic tones and the power of
harmonic tones. It is equivalent to the noise-to-harmonics ratio if the
nonharmonic component of the signal is subharmonics dominated. Stronger
subharmonics yield higher SHR measures and are expected to increase the
\(f_o\) detection errors. As such, studying the SHRs of the signal
intervals with ACF subharmonic errors identifies the sensitivities of
the \(f_o\) detectors to the subharmonic strength. The SHR was computed
by
\begin{equation}{SHR = \frac{\sum\limits_{k \in \mathcal{K}_s} S_{xx} \left(k\hat{f}_o\right)}{\sum\limits_{k\in \mathcal{K}_h} S_{xx}\left(k\hat{f}_o\right)},
}\end{equation} where
\(\mathcal{K}_h = \{ p M: p = 1, 2, \ldots, \lfloor K(\hat{f}_o)/M \rfloor \}\)
is the set of the harmonic multipliers, and
\(\mathcal{K}_s = \{1,2,\ldots,K(\hat{f}_o)\} \setminus \mathcal{K}_h\)
is the set of the subharmonic multipliers. Here, \(\hat{f}_o\) is the
ACF estimate, and \(M (\approx f_o^*/\hat{f}_o)\) is its associated
subharmonic period. The signal frame shown in Fig. \ref{fig:msub_demo}
registered --6.8 dB SHR.

\section{Results and Discussion}\label{sec:res_n_disc}

\begin{figure*}
\centering
\includegraphics[keepaspectratio]{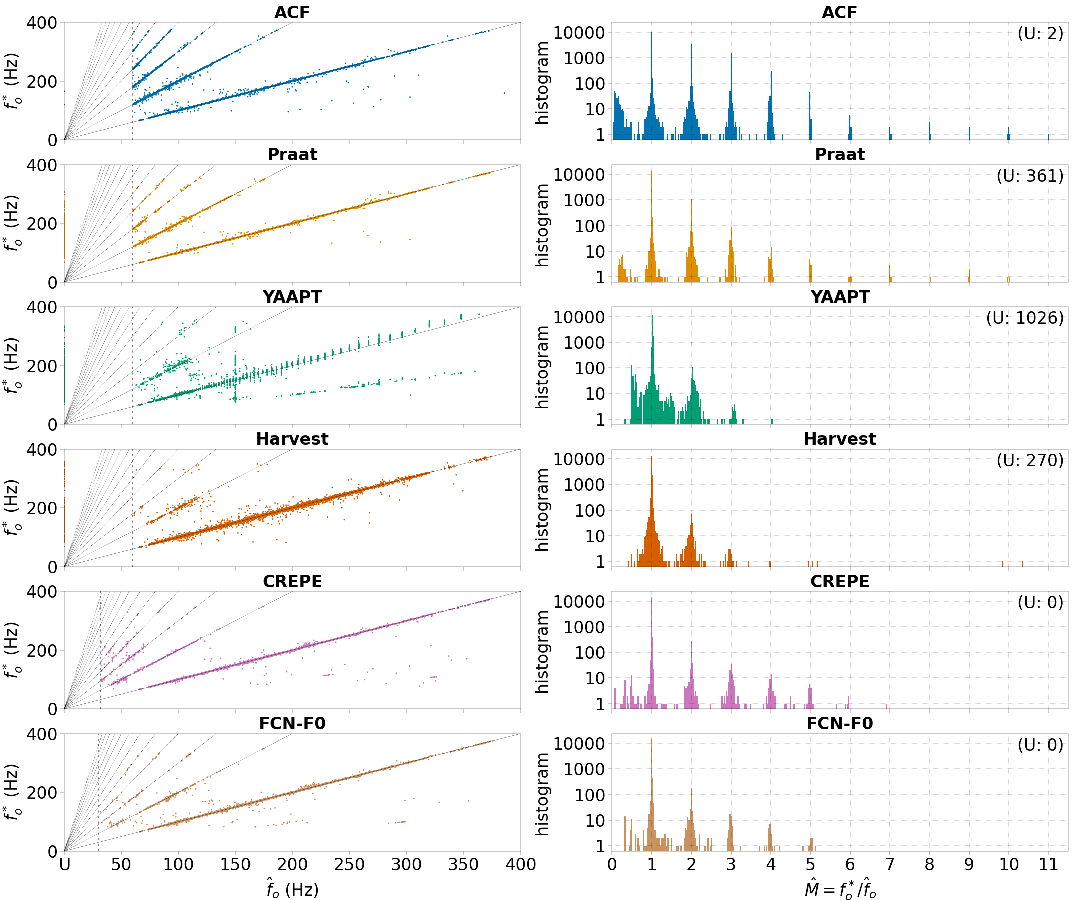}
\caption{(color online) (left column) Scatter plots of estimated
\(\hat{f}_o\) (\(x\)-axis) vs.~manually selected truth \(f_o^*\)
(\(y\)-axis) of the six \(f_o\) estimators under study. Diagonal grid
lines indicate the mapping of correct and subharmonic estimates (\(M=\)
1 to 7). Vertical dotted lines indicate the minimum \(f_o\) imposed by
each estimator. The samples at \(\hat{f}_o=0\) (U) indicate the
intervals which were marked as unvoiced by the \(f_o\) estimator. (right
column) Histograms of the period elongation factor of the estimates,
\(\hat{M} = f_o^*/\hat{f}_o\). (15941 samples per
plot)}\label{fig:scatters}
\end{figure*}

A scatter plot of \(f_o^*\) vs.~\(\hat{f}_o\) for each \(f_o\) estimator
is shown in Fig. \ref{fig:scatters} with diagonal grid lines indicating
the correct and subharmonic mappings. The cases with abnormally high
\(f_o^*\) (shown in Fig. \ref{fig:fo_dist}) or \(\hat{f}_o\) are not
displayed because they are scarce (less than 2\% of all signal
intervals) and are an atypical response to the instruction of producing
voice at normal pitch. The top scatter plot reveals the hypersensitivity
of the ACF to subharmonics. Fig. \ref{fig:scatters} also shows the
period elongation factor \(\hat{M} = f_o^*/\hat{f}_o\) of each
estimator. Integer \(\hat{M}\)'s correspond to the subharmonic periods.
Subharmonics with periods from 2 to 6 are observable in the scatter
plots with the highest period of 11. The Praat result indicates that its
Viterbi postprocessor eliminated a number of high subharmonic period
cases though many still remain.

YAAPT and Harvest, which are both more complex than Praat, are visibly
less sensitive to subharmonics. Not many of their estimates follow the
subharmonic grid lines except for the period doubling. Instead, their
errors have different tendencies (which lead to more ``other error''
labels). YAAPT tends to report erroneous \(\hat{f}_o = 150\) Hz, and its
granularity and bias are apparent at high \(f_o\). It is also
susceptible to the frequency-doubling error with its results forming the
line \(\hat{f}_o = 2f_o^*\). The estimates of Harvest, on the other
hand, show higher variance along the correct mapping.

The mappings of the deep-learning solutions, CREPE and FCN-F0, are
placid compared to the others. While their subharmonic errors are
apparent, they are much more subdued than Praat. FCN-F0 visibly has less
estimates with higher subharmonic errors than CREPE. The results do not
indicate any other obvious tendency of their estimation errors.

\begin{figure*}
\centering
\includegraphics[keepaspectratio]{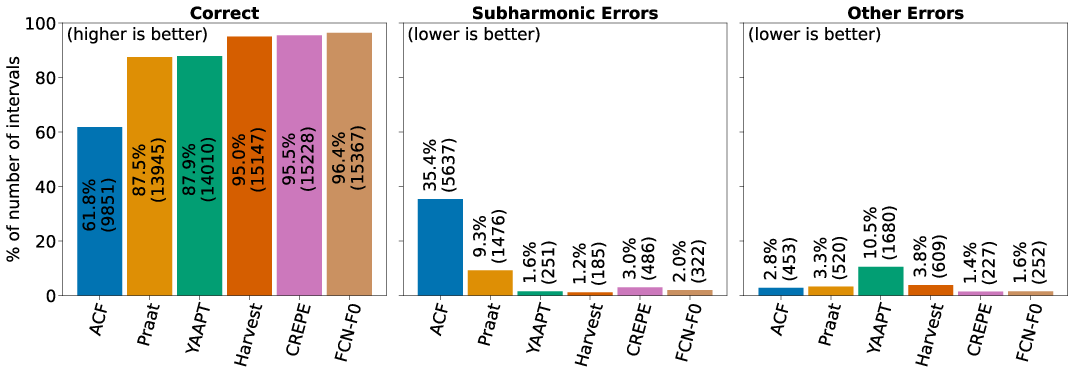}
\caption{(color online) Quality-of-estimate outcomes: Correct \(f_o\)
estimation rates, subharmonics error rates, and other error rates of the
\(f_o\) estimators under study. Numbers in parentheses are the number of
intervals out of 15941 intervals.}\label{fig:per_frame_rates}
\end{figure*}

Next, the per-interval rates of correct and erroneous estimations are
presented in Fig. \ref{fig:per_frame_rates}. FCN-F0 marked the highest
accuracy with a 96\% success rate, closely followed by CREPE and Harvest
at 95\%. YAAPT and Praat (88\%) were below the top three and trailed by
ACF (62\%).

The types of the estimation errors follow the observation of
Fig. \ref{fig:scatters}. The ACF and Praat are dominated by subharmonic
errors though the Praat postprocessor reduced the error by 74\%. The
rate of the other types of errors is low for both ACF and Praat. YAAPT
produced more other errors (11\%) in exchange for minimal numbers of the
subharmonic errors (1.6\%). Harvest likewise produced more other errors
than subharmonic errors but only the third as many other errors than
YAAPT. Finally, the deep-learning models produced the most evenly
balanced ratio of subharmonics to other errors. The FCN-F0 model appears
to be the best solution not only for the highest accuracy but reduces
the subharmonic errors (34\% less than CREPE) and achieved the best
balance between subharmonic and other errors (2.0\% vs.~1.6\%).

As predicted, the most simplistic ACF estimator failed most often, and
the vast majority of its failures were attributed to the subharmonic
errors (35.4\% vs.~2.8\% other). Additional insights may be gained by
studying how the signal intervals were evaluated differently by the
other estimators compared to ACF.

\begin{figure*}
\centering
\includegraphics[keepaspectratio]{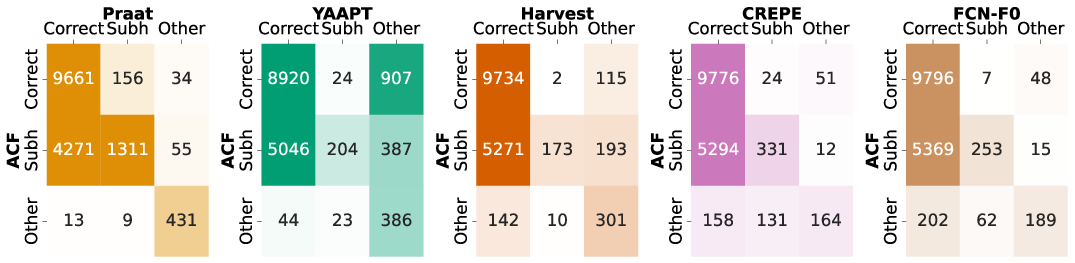}
\caption{(color online) Contingency tables of the outcomes of the ACF
estimator vs.~the other five \(f_o\) estimators. Numbers indicate the
number of intervals out of 15941
intervals.}\label{fig:per_frame_confmtx}
\end{figure*}

Fig. \ref{fig:per_frame_confmtx} shows the contingency tables,
illustrating how the ACF outcomes differ from the others' outcomes. An
ideally improved estimator maintains all the ACF's correct estimations
while it corrects as many of the cases of which ACF failed. The FCN-F0
model came the closest to this ideal. It reduced both subharmonic and
other error counts while keeping most of the ACF's correct decisions.
The CREPE model also came close; however, its resolution for the other
errors was notably inferior to FCN-F0 (reclassified 29\% of ACF's
other-error cases to subharmonic errors).

Both deep-learning models maintained most of ACF's correct estimates
(CREPE only turned over 0.8\% and FCN-F0 0.6\%). Harvest lost 1.1\% to
other errors, mostly due to its high estimation variance while Praat
switched 1.6\% to subharmonic errors due to the Viterbi algorithm's
breakdown on recordings, which comprise majority subharmonic. The
Viterbi algorithm may incorrectly toggle the correct \(f_o\) estimates
to \(f_o/M\) when subharmonic frames outnumber harmonic (or weakly
subharmonic) frames in a recording. Of YAAPT's 907 turnovers to the
other error, it lost 38\% to the 150-Hz error and another 42\% to the
frequency-doubling error.

On the other end of the tables, Harvest, CREPE, and FCN-F0 were able to
turn the ACF's other error cases to match the annotated (31\%, 35\%, and
46\% turnover rates, respectively). These are the cases, in which the
non-harmonic component presents a stronger peak of the autocorrelation
function than the harmonic component. Praat and YAAPT were generally
incapable of handling such cases.

Finally, the ACF's subharmonic error cases, which are the cases with
suspected presence of subharmonics, are either resolved and remained as
subharmonic error by FCN-F0, CREPE, and Praat. YAAPT and Harvest both
switched more than half of their remaining errors to the other error
category. The majority of the Harvest's switches were either due to high
variance or unvoiced decisions. Interestingly, none of the YAAPT's
switches was due to unvoiced decisions, and no \(\hat{f}_o\) patterns
were visually apparent.

\begin{figure}
\centering
\includegraphics[keepaspectratio]{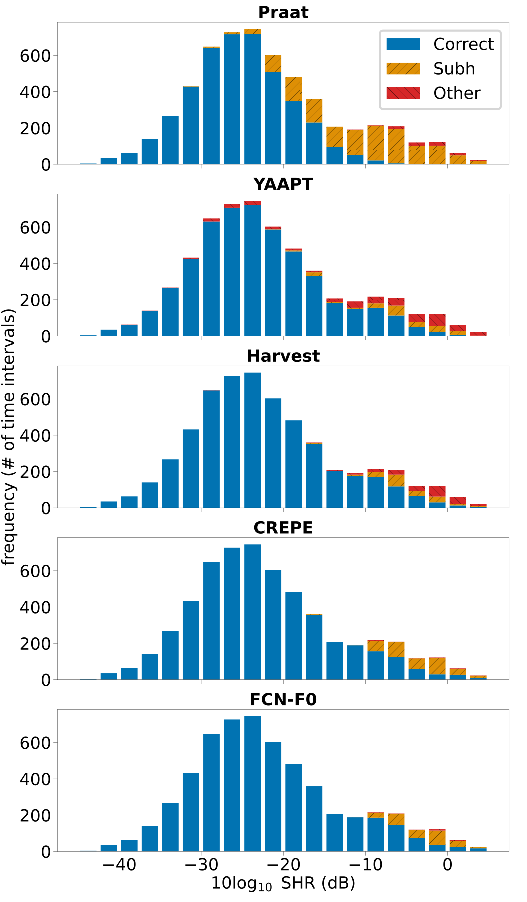}
\caption{(color online) Conditional histogram of the SHRs of the signal
intervals in which the ACF registered subharmonic errors. Partitioned
histograms are shown to illustrate how the other five estimators handled
these intervals: correct, subharmonic (subh) error, or other
error.}\label{fig:shr_hist}
\end{figure}

Further insight for the handling of subharmonic error cases can be
gained by analyzing the estimators' outcomes with the SHR measurements
as shown in Fig. \ref{fig:shr_hist}. All the histograms show the same
overall shape, i.e., the occurrences of ACF subharmonic errors as a
function of the SHR. It is apparent that there are two peaks in this
distribution: the main peak at --25 dB and the secondary peak at --10
dB. Based on a few spectral samplings of the cases with low SHR, we
postulate that the main peak represents the incidental estimation errors
by ACF while the secondary peak represents the signal intervals with
subharmonics. All other \(f_o\) detectors were able to fix most of the
ACF's incidental errors in the main peak. Praat's inability to correct
most of the subharmonic errors in the second peak indicates that Praat's
postprocessor only excels in correcting the incidental ACF errors, and
its reported subharmonic error rate (9.3\% in
Fig. \ref{fig:per_frame_rates}) is a reasonably tight lower bound for
the portion of the signal intervals with subharmonics. The true number
of subharmonic intervals would be higher because the imposed minimum
\(f_o\) limit (60 Hz) forces Praat and ACF to ignore \(f_o/M\)
candidates below the limit, steering them away from committing the
subharmonic errors. This is most prevalent for the male voices with
period-doubling subharmonics.

Note that SHR = --10 dB (0.1) is roughly equivalent to period-2
subharmonic amplitude modulation with 10\% modulation
extent\citep{herbst2021}, and Bergan and Titze\citep{bergan2001} found
that amplitude modulation with 10\% modulation extent is a lower bound
for period-doubling subharmonic voicing to produce an audible effect.
FCN-F0, CREPE, and Harvest resolved virtually all cases with SHR
\textless{} --10 dB, and they were also able to handle a sizable number
of cases with SHR \textgreater{} --10 dB. Among the 755 recording
intervals with the SHR \textgreater{} --10 dB, Praat estimated 3.4\% of
them correct, YAAPT 46.0\%, Harvest 54.3\%, CREPE 53.4\%, and FCN-F0
63.7\%.

The report by Vaysse et al.\citep{vaysse2022} that YAAPT and FCN-F0 can
handle the subharmonic intervals over Praat is confirmed in
Fig. \ref{fig:shr_hist}. YAAPT and FCN-F0 show that they registered a
number of correct estimates well into the \textgreater{} --10-dB SHR
region. There is also a notable discrepancy between these current
findings and Vaysse et al.~They found that NDF (which is a predecessor
of Harvest\footnote{Both NDF and Harvest were initially considered for
  this study, and only Harvest is presented in this paper because they
  are conceptually similar and Harvest decisively outperformed NDF
  (95.0\% vs.~88.4\% in their accuracies).}) had the lowest
halving-\(f_o\) errors, lower than FCN-F0 by more than a percentage
point for the head-and-neck cancer group. This is in contrary to the
current study in which FCN-F0 outperformed Harvest by 9\% for the
accuracy for the cases with SHR \textgreater{} --10 dB. A likely reason
for these discrepancies is the target recording type. The presented
results are for sustained /a/ recordings while Vaysse et al.~analyzed
running speech signals. Sustained vowel phonation is naturally more
consistent than connected speech. The performance of deep-learning
models depends on the data they were trained on, and the combination of
unfamiliar intonation patterns and unfamiliarity to pathological voicing
may have caused these models to have difficulty processing running
speech samples.

There is an interesting parallel between the model-based Harvest
estimator and the data-based CREPE and FCN-F0 estimators. All three
estimators make use of frequency binning. Harvest uses a filterbank as
the first step while the deep-learning models produce a likelihood
measure for each bin. It could be that the deep-learning models are
trained to assimilate the internal working of Harvest, and FCN-F0
surpassed it for sustained vowel application.

A limitation of this study is the uncertainty in the manually determined
\(f_o\) truths. The nearly periodic nature of subharmonic voicing made
the establishment of the truths challenging for the voices with
sustained strong subharmonics. Most notably, a female voice with strong
subharmonics could be mistaken for a normal male voice. Despite careful
annotation effort, it is possible for a small number of such errors to
be still present in the truths. This could slightly change the reported
accuracies of the \(f_o\) detectors, but the key finding remains valid:
None of the \(f_o\) detectors could reliably handle strong subharmonics
(SHR \textgreater{} --3 dB). Use of glottal inverse filtering or a
dataset with simultaneous source observations (e.g., high-speed
videoendoscopy or electroglottography) could minimize this type of
errors.

\section{Concluding Remarks}\label{concluding-remarks}

Deep learning appears to be the best approach to estimate the
fundamental frequency of pathological sustained voice, especially with
its encouraging results in handling the voice signals with subharmonics.
Detecting the speaking fundamental frequency of subharmonic voicing is a
challenging process to automate as most algorithms are designed to
estimate the pure mathematical definition of the fundamental frequency
rather than the speaking fundamental frequency which is tied to the
overall glottal behavior (i.e., opening and closing). Deep learning
models are generally free from such algorithmic assumptions as their
architectures are generic, and their model coefficients are trained by
annotated data. These models can, therefore, utilize more features than
just periodicity to reach its conclusion. For example, relative
amplitudes and phases of the harmonics may provide additional
information.

In this study, the two deep-learning models, CREPE and FCN-F0,
outperformed other estimators in the quality-of-estimate metric. They
were in agreement with the manually annotated truths for more than 95\%
of the cases, slightly exceeding the performance of Harvest, the leading
non-data-driven estimator. Meanwhile, CREPE and FCN-F0's \(f_o\)
estimates are visibly less variable than those of Harvest. The
deep-learning models also demonstrated their consistent ability to
handle weak subharmonics with the SHRs under --10 dB, and their
subharmonic errors are compartmentalized only to the high SHR cases.
Achieving this degree of performance without being trained with
subharmonic voice samples is quite remarkable. Retraining these models
with subharmonic voice samples may bring further performance
improvements for the high SHR cases.

Accurate speaking fundamental frequency estimate is crucial to the
advancement of acoustic voice analysis, especially in clinical use. Deep
learning is a promising tool to address the current shortcomings with
the handling of subharmonic voicing.


\end{document}